\documentstyle[11pt,epsfig]{article}
\begin{document}
\baselineskip 18.0pt
\def\oneskip{\vskip\baselineskip}
\def\xr#1{\parindent=0.0cm\hangindent=1cm\hangafter=1\indent#1\par}
\def\la{\raise.5ex\hbox{$<$}\kern-.8em\lower 1mm\hbox{$\sim$}}
\def\ma{\raise.5ex\hbox{$>$}\kern-.8em\lower 1mm\hbox{$\sim$}}
\def\ea{\it et al. \rm}
\def\am{$^{\prime}$\ }
\def\as{$^{\prime\prime}$\ }
\def\msol{M$_{\odot}$ }
\def\la{\raise.5ex\hbox{$<$}\kern-.8em\lower 1mm\hbox{$\sim$}}
\def\ma{\raise.5ex\hbox{$>$}\kern-.8em\lower 1mm\hbox{$\sim$}}
\def\ea{\it et al. \rm}
\def\am{$^{\prime}$\ }
\def\as{$^{\prime\prime}$\ }
\def\msol{M$_{\odot}$ }
\def\kms{$\rm km\, s^{-1}$\ }
\def\cm3{$\rm cm^{-3}$}
\def\Ts{$T_{\rm *}$}
\def\Vs{$V_{\rm s}$}
\def\n0{$n_{\rm 0}$}
\def\B0{$B_{\rm 0}$}
\def\Fn{$F_{\nu}$}
\def\ne{$n_{\rm e}$}
\def\Te{$T_{\rm e}$}
\def\Tgr{$\rm T_{gr}$}
\def\Tgas{$\rm T_{gas}$}
\def\Ec{$\rm E_{c}$}
\def\erg{$\rm erg \, cm^{-2} \, s^{-1}$}
\def\ergs{$\rm erg\, s^{-1}$}
\def\hii{H{\sc ii}}
\def\Hb{H$\beta$}
\def\oi{O\,{\sc i}}
\def\oiii{O\,{\sc iii}}
\def\oiv{O\,{\sc iv}}
\def\neii{Ne\,{\sc ii}}
\def\neiii{Ne\,{\sc iii}}
\def\siii{S\,{\sc iii}}
\def\slii{Si\,{\sc ii}}
\def\cii{C\,{\sc ii}}
\def\kms{$\rm km \, s^{-1}$}
\def\cm3{$\rm cm^{-3}$}
\def\Ts{$\rm T_{*}$}
\def\Vs{$\rm V_{s}$}
\def\n0{$\rm n_{0}$}
\def\B0{$\rm B_{0}$}
\def\Fh{$\rm F_{H}$}
\def\Fn{$\rm F_{\nu}$}
\def\ne{$\rm n_{e}$}
\def\Te{$\rm T_{e}$}
\def\Hb{H$\beta$}
\def\Tgr{$\rm T_{gr}$}
\def\Tgas{$\rm T_{gas}$}
\def\Ec{$\rm E_{c}$}
\def\N{$\it N$}
\def\ff{$\it ff$}
\def\sf{$\it \sqrt{ff}$}
\def\erg{$\rm erg \, cm^{-2} \, s^{-1}$}
\def\ROIII{$\rm R_{OIII}$}
\bigskip
\bigskip

\centerline{\large{\bf The interpretation of the emission spectra of the Seyfert 2 
galaxy NGC 7130.}} 

\centerline{\large{\bf Determination of the AGN and starburst contributions.}}

\bigskip

\bigskip

\bigskip

\centerline{M. Contini$^{1,3}$, M. Radovich$^{4}$, P. Rafanelli$^{2}$,
and G.M. Richter$^{3}$}

\bigskip

\bigskip

\bigskip
   
$^1$ School of Physics \& Astronomy, Tel Aviv University, 
69978 Tel Aviv, Israel

$^2$ University of Padova, Department of Astronomy, Vicolo dell'Osservatorio,
5, I-35122 Padova, Italy

$^3$ Astrophysikalisches Institut Potsdam, An der Sternwarte, 16, D-14482
Potsdam, Germany

$^4$ Osservatorio Astronomico di Capodimonte, via Moiariello 16, I-80131, Napoli, 
Italy

\bigskip

\bigskip

\bigskip

\bigskip

Running title : Spectra of NGC 7130

\bigskip

\bigskip

\bigskip

subject headings : galaxies: individual: NGC 7130 
 - galaxies: AGN - 
galaxies: starburst -  galaxies - line: models

\newpage

\section*{Abstract}

Detailed modeling of the different regions of NGC 7130 is presented,
accounting for its composite nature of an AGN and a starburst galaxy.
Shock waves,  created by stellar winds
from hot massive stars and by supernova ejecta, are evident
in the  continuum 
and line spectra emitted from the clouds. 
Therefore, the SUMA code, which accounts consistently for
photoionization and shocks is adopted in model calculation.
The results show  that  the nuclear region is dominated by
 gas  ionized by a  power-law radiation flux from
the active center (AC).
 High  velocity (\Vs = 1000 \kms) clouds, which account for
the broad FWHM component of the line profile, are found close to the AC,
and are characterized by a high dust-to-gas ratio ($>$ 10$^{-12}$), while 
the dust-to-gas ratio is about $10^{-14}$ throughout the galaxy. 
Massive stars with temperature of 5-7 $10^{4}$ K  photo-ionize and heat the gas in
 the outer regions and an old star population (\Ts = 3000 K) background radiation 
contributes  to the fit 
of the continuum in the optical - near IR range.
The AGN starburst connection is discussed on the basis of model results,
considering, particularly, the distribution of densities and velocities 
throughout the galaxy.

\newpage

\section{Introduction}

NGC 7130 is  a  remarkable prototype of "composite"
galaxy, i.e. a galaxy containing an Active Galactic Nucleus (AGN) 
surrounded by starburst (SB) regions both in the nuclear and 
circumnuclear regions.
The presence of a Seyfert nucleus and of an energetic SB are 
evident from the optical and near-IR properties of the galaxy.

NGC 7130 (z=0.016, morphological type Sa pec) was classified as a Seyfert 2 galaxy 
by Phillips, Charles, \& Baldwin (1983).
The object was observed in details by Shields \& Filippenko (1990). 
They report on spectra of nuclear and off-nuclear regions extracted from one 
long slit spectrum taken at P.A. = $11.5^o$. 
The nuclear and circumnuclear regions of NGC 7130 were recently observed by
Radovich et al. (1997 hereafter RRBR) 
and emission line ratios were measured in the different regions (see Figure 1).
Two long-slit spectra   were taken at La Silla on
November 1996 with the ESO 2.2m telescope + EFOSC2 and a 2" wide long-slit
oriented at position angles $90^o$ and $160^o$ through the nucleus of
  the galaxy.
Recently, HST images and UV (GHRS) spectra plus ground based optical
spectra were presented by Gonz\'alez Delgado et al. (1998) investigating the
AGN-SB connection.

Emission lines in all the observed regions in NGC 7130 have a complicate 
profile (RRBR) indicating that different conditions coexist.
Particularly, the line profiles observed in the inner $2^"$ show a complex 
structure with two different main components: a broad  line of about 1000 \kms 
and a narrow one of $\sim$ 200 \kms in the forbidden and permitted lines.
These FWHM are characteristic of both AGN  and SB line profiles.
Velocities  $\geq$ 1000 \kms were recently observed in the NLR
of AGN (e.g. NGC 4151, Hutchings et al. 1999). 

Indeed shock waves must form in the tormented regions containing the SB. 
Shocks are also present in the narrow line-emission
region (NLR) of AGN, due to collision of high velocity clouds.
The presence of shocks in AGN has been generally  confirmed  both by the
observations and model calculations.
In the particular case of NGC 7130 shocks can  quite likely be predicted either 
by the outflow of gas due to stellar winds from hot massive stars 
and/or by supernova ejecta.

The analysis of AGN has drastically improved in the
last years as soon as observations in the different regions of a single galaxy
became available.
The spectra of NGC 7130 have been previously  analyzed
by RRBR comparing the most indicative observed line ratios with model 
results. The code CLOUDY (Ferland 1996) which accounts for pure photoionization 
effects was used.
Notice, however, that the  spectral lines observed by RRBR  correspond to 
ionization levels within the III one. These lines are generally strong when the 
emitting gas is ionized by  radiation. The  shock, on the other
hand, affects lines from higher levels,  because the downstream emitting gas  can 
reach higher temperatures.
Moreover, the comparison of the observed line ratios in the Baldwin,
Phillips, \&Terlevich (1981, Fig. 5) diagrams, shows that NGC 7130 spectra 
are typical of regions ionized both by photoionization and shocks.
 
An important step in the  modeling  process is to    explain the  spectra
in the different regions of the galaxy in order to determine the nature of
the photo-ionizing source: AGN or/and SB  . 
The models are  constrained   by the agreement of  line and continuum
spectra, leading   to precious informations
about the structure of the galaxy  (e.g. Contini, Prieto \& Viegas 1998).

In this work  we would  like, therefore, to treat NGC 7130 by considering
all the available lines in single spectra and the continuum spectral
energy distribution (SED), including the effect of the shock. 
Actually, a more complete picture 
of the galaxy  emerges from the dynamical analysis of its different regions.
The code SUMA which accounts consistently 
for both the ionization by an external source and for the shock
is adopted.
A complete description of  SUMA is given by  Contini \& Viegas (2001,
and references therein).
The relative importance of the two ionizing sources (AGN and SB) is 
investigated on the basis of  RRBR observations. 

Finally, the results of  the spatial distribution of the velocity field, 
of  the gas  density, of the intensity of radiation from the AC, and of 
the temperature of the stars throughout the galaxy,
can be used to  discuss the AGN-SB connection.

Models are presented in \S 2.
The line spectra  both from the narrow line component and  the broad line
component are calculated and compared to observations in the 
different regions of the galaxy  in \S 3.
In \S 4 we discuss the observed continuum SED and the 
contributions from individual single-cloud models; the fit  by a multi-cloud
averaged model is then presented. 
%
The structure of NGC 7130 is discussed in \S 5 and conclusions follow in \S 6.

\section{The models}

 The  general model is based on previous works by
 Contini et al. (2001), Viegas \& Contini (1994), Viegas-Aldrovandi \& Contini (1989), 
and references therein, regarding the NLR of AGN:   
gaseous and dusty clouds emitting the line and continuum spectra move  either outward 
(see Fig. 2) from the active center (AC) or inward toward the AC.
A grid of models is calculated for NGC 7130 both in the case of a power-law (pl)
radiation from the AC and of black body (bb)  from stars; however,
only the best fitting models are shown in the tables.
The models  account consistently for the shock, for the photoionizing radiation
flux from an external source,
and for diffuse radiation emitted by hot slabs of gas within the clouds.

The input parameters are: \Vs, the shock velocity, \n0, the preshock density, 
\B0, the magnetic field, \Fn,  the 
radiation flux from the AC in units of number of photons $\rm cm^{-2} \, s^{-1}
\, eV^{-1}$
at 1 Ryd, and the spectral index $\alpha$, in the power-law case; 
U,  the ionization parameter and \Ts, the color temperature of the stars, in the
black--body case;
D is the geometrical thickness of the emitting clumps and d/g, the dust-to-gas 
ratio by number.  
Abundances of He, C, N, O, Ne, Mg, Si, S, CL, and Fe relative  to H 
are basically  taken from Allen (1973), and eventual changes are discussed 
in the text.

The ranges of the input parameters were chosen  with the following
criteria, however, they were then adjusted in order to achieve the
best fit to the data:

1) The shock velocity is deduced roughly from the FWHM of the line profiles, but
it is not always necessarily the FWHM.

2) The intensity of the radiation flux from the AC is  evaluated
from  Viegas-Aldrovandi \& Contini ( 1989) and Contini \& Viegas (2001), 
comparing the [O III] 5007/\Hb,  [Ne III] 3869/\Hb,
~and He II 4686/ \Hb ~line ratios.

3) The density of the emitting gas is roughly deduced from the 
[S II] 6716/6730 line ratios.

4) The temperature of the stars and  the ionization parameter are determined
directly from the fit of model calculations.

5) The geometrical thickness of the clouds   is a crucial parameter
which can be hardly deduced from the observations. 
Indeed, in the turbulent regime created by shocks, R-T and K-H instabilities cause
fragmentation of matter which leads to clouds with very different thickness 
coexisting in the same region. 

From the modeling point of view, once  the physical conditions of the 
emitting gas, and the photo-ionizing type and intensity  are determined
by the appropriated line ratios, the geometrical thickness of the cloud
is deduced  from the  ratio between high and low ionization level lines.
In fact, larger clouds will contain a larger volume of gas at a relatively low
temperature, leading to stronger low ionization level lines.
The choice of D, within the range  of the observational evidence,
is then constrained  by the best fit of a large number of  line ratios. 

The clouds are matter-bound or 
radiation-bound depending on the geometrical thickness, as well as on 
\Fn ~ and \Vs.

6) The dust-to-gas ratio is determined by the SED of the continuum in the IR
range.

7) For all models a magnetic field \B0 = $10^{-4}$ gauss is adopted.

8) A spectral index $\alpha$ = 1.5 characterizes the power-law radiation.
Different values are tested during the fitting process of calculated to observed 
spectra, considering particularly that the HeII/\Hb
~line ratio is sensibly affected by the choice of $\alpha$. 
The adopted  value of 1.5  is generally in agreement with Seyfert 2 
galaxy spectra.

We have run models considering both
inflow and outflow of gas, with the  aim of explaining the spectra of NGC 7130
and providing, moreover, some  information about the strongest line ratios,
which could be  of general use.

 Fig. 3 compares the observed [OI] 6300+/\Hb ~ versus [OIII]5007+/\Hb 
~with the values calculated within a 
reasonable   range of the most significant input parameters in 
the inflow case.  The lines are strong and the observational error
small enough to exclude such models for NGC 7130, on the basis of an insufficient
agreement. Particularly, in the range of the observed [OIII]/\Hb, the calculated
[OI]/\Hb ~ are too low for models  corresponding to \n0=50-100 \cm3 and too
high for \n0=200 \cm3.
For all models \Vs = 200 \kms is adopted, as suggested by the FWHM of the line profiles.

On the other hand, models  referring to an outflow better explain the line
ratios and will be adopted to model the galaxy. 
The presence of outflowing gas in the nuclear regions of NGC 7130 is 
confirmed  (Gonz\'alez Delgado et al. 1998) by the fact that interstellar 
absorption lines as SiII 1260 and SiII 1526 
are blueshifted by 720 \kms with respect to the systemic velocity. 
A more indirect evidence for outflowing gas 
in the outer regions ($R \le 16$ kpc) has been provided by Levenson et al. (2001): 
they showed that the soft X-ray emission in NGC 7130 is dominated by thermal 
emission whose origin is the outflowing superwind in a starburst. 

In the outflow case a  shock front forms on the outer edge of the
clouds.
The radiation flux from the  AC reaches the inner edge of the clouds opposite
to the shock front.
  A different situation occurs for the SB regions, which are  distributed 
in both  the
nuclear and the circumnuclear regions of the galaxy. 
Radiation from stars reaches either the inner or the outer edge of the clouds
depending on their location in the galaxy. More particularly, 
  black body radiation  from  hot stars located in the nuclear region
 reaches the inner edge of the clouds opposite to the shock front, whereas,
radiation from hot stars located in the circumnuclear
 regions reaches the very shock front edge of the clouds  (Fig. 2). 

\section{Comparison of calculated with observed spectra}

The comparison of the observed line ratios I($\lambda)$/I(\Hb) with model 
calculations is presented in  Tables 1 and 2 for the nuclear region
and in Tables 3 and 4 for the circumnuclear region; also given are 
 the distance of the center of each region  from the peak in the continuum  
($r_c$) and its size  ($\Delta r$).  
The + indicates that both components of the doublet are summed up.
 The ionizing continuum is a power-law for models PL1-PL3, a black-body
for models BB1-BB9. 
In the latter case the radiation source is not necessarily in the very center
of the galaxy. 
The input parameters of the models appear in the  bottom of the tables. 
The best fitting models were generally selected among those 
corresponding to  stars  external to the emitting cloud region.

\subsection{The nuclear region}

Three regions (N1, N2, and N3) were observed within  
$\sim 2^"$ from the nucleus (see Figure 1) at P.A. = $90^o$ and at P.A. = $160^o$. 

\subsubsection{ Regions centered on the nucleus at P.A. = $90^o$}


Table 1 refers to the nuclear region  observed at P.A.=$90^{o}$. 
The spectrum observed from region N1 is well fitted by  model PL1 with a 
shock velocity \Vs = 240 \kms, in good
agreement with the observed FWHM of the line profiles. The preshock density
is rather low, \n0 $\sim$ 100 \cm3, also suggested by
the [S II] 6716/6730 line ratio. The [O I] 6300/ \Hb
 ~and [S II]/ \Hb ~line ratios are, however, better fitted by  model PL2 
also  characterized by a
power-law but with a slightly lower \Vs ~and a lower preshock density, \n0.
Model BB2 which accounts for bb radiation from the  stars also explains some 
of the line ratios observed in region N1. 
The calculated absolute \Hb ~flux of model BB2 is lower by a factor  of $>$ 100
than that of PL1 and PL2 which are power-law dominated models, therefore, 
the effect of  star radiation in the  spectrum of N1  could be felt only adopting 
a high weight of the model BB2 in the averaged sum of PL1, PL2, and 
BB2 spectra. 

The spectrum observed from region N2 is  well fitted  by model PL3 
  with a preshock density \n0 = 200 \cm3; black-body models (e.g. BB3) are 
not able to simultaneously reproduce the high [O I] 6300+/\Hb ($\sim 0.4$) and
 [O III] 5007+/\Hb ($\sim 8$) ratios. 
The HeII/\Hb ~from N2 is poorly fitted by model PL3.
Notice, however, that Gonz\'alez Delgado et al (1988) claim that the HeII 4686
 line cannot be properly estimated due to the absorption features close
to 4686 \AA. Consequently the flux and the FWHM cannot be estimated to
better than within a factor of 2. Moreover, 
Gonz\'alez Delgado et al find that \Hb ~is resolved in two components: the narrow
one with $\sim$ 300 \kms ~and the broad one with $\sim$ 960 \kms. Therefore,
both the narrow and the broad component
must be accounted for (\S 5).
The radiation flux from the AC, \Fn, adopted for model PL3  is higher by more
than a factor of 20 than
those adopted in explaining  the N1 spectrum (PL1, PL2). This
indicates that either region N2 is closer to the AC or that the cover factor is
lower. Densities
are higher than in the other positions, in agreement with [S II] 6716/6730 $<$
1.
We recall that clouds in the same region of the galaxy  can show very
different values of D (see \S 2).
Particularly, in  region N2, 
[OIII]/\Hb ~is higher  than in N1.  N2 is closer to
the active center, therefore reached by  a stronger  \Fn ~which 
 maintains the gas hot and ionized in a large region of the cloud.
However, the [OI]/\Hb ~line ratio is higher in N2 than in N1.
This indicates that a large region of cold gas is  present
inside the clouds.
Consequently, the geometrical thickness of the emitting clouds in N2
is larger than in N1. We do not refer to the other neutral line, [NI],   
because the critical density for collisional deexcitation is
very low ($<$ 1000 \cm3). As densities are higher in
N2 than in N1  the effect of D is washed out.

The spectrum observed in region N3 is rather poor in number of lines and is
explained only by  model BB4
characterized by black body radiation from stars with a temperature of 5 $10^4$
K.
Power-law models do not fit because the calculated low [O III] 5007/ \Hb ~line
ratio implies higher than observed [O I]/ \Hb ~and [O II]/ \Hb ~line ratios.

Finally, the geometrical thickness of the clouds  in the  N1 and N3 regions
observed at position  P.A. = $90^o$ is not larger than 0.2 pc, while in the N2 
region the emitting gas corresponds to  clumps  of  $\sim$ 1 pc.

\subsubsection{Regions centered on the nucleus at P.A. = $160^o$}


The spectra observed at P.A. = $160^o$ are compared with model calculations in
Table 2. 
The internal region N2 shows  the same conditions
as found at  P.A. = $90^o$. N1 and N3 spectra, on the other hand, are dominated
(besides the effect
of the shock) by stars with a temperature of about 7.5 $10^{4}$ K 
(Table 2, model BB5).

 The observed [Ne III]/ \Hb ~is extremely
high in  both the N1 and N3 regions, 
much higher than  that indicated by any diagnostic diagram for 
[O III] 5007 / \Hb $\leq$ 10. 
A  satisfactory fit is obtained adopting Ne/H relative abundance of 3.2
$10^{-4}$
which is 4 times higher than cosmic.  Even the solar value of Ne/H = 1.17
$10^{-4}$
given by Ferguson et al (1997) is not high enough and 
a double value should be
necessary to fit the line ratio. Such high values are  given by Osterbrock
(1965) and Burbridge et al (1963) for planetary nebulae, and are  generally
accompanied by a higher O/H.  It is however possible that, due to the low signal
to noise ratio in these regions, the template was not properly subtracted 
and as a consequence the intensity of \Hb ~was underestimated.

\subsubsection{Morphological structure of the nuclear region}

The distance of the emitting nebula from the center of the galaxy
can be roughly deduced from the comparison of observed and calculated \Hb
~absolute fluxes.
We start with region N2 which is closer to the nucleus and adopt model PL3  as
the
best fitting model to the line spectrum. We refer to P.A. = $ 90^o$.
The geometrical thickness of the
typical cloud in region N2 is D = 0.77 pc. The N2 region is $\sim 2"$ wide
corresponding to 620 pc. The number of clouds \N ~is then $\sim $ 800 \ff, where
\ff ~is the 
filling factor, assuming that all clouds are similar.
 If R(N2) is the distance of N2 from the nucleus, then
$\rm d^2 $ $\rm H\beta_{obs}$ = $\rm R(N2)^2$ $\rm H\beta_{calc}$ 800 \ff.
If the distance from Earth is d = 64 Mpc, and  $\rm H\beta_{obs}$ = 4.9
$10^{-13}$
\erg (RRBR), R(N2) = 2 / \sf ~pc. 
Adopting \ff $\leq$ 1 for the compact central knot, R(N2) $>$ 2 pc in good 
agreement with the measures of Shields \& Filippenko (1990).

Similar calculations for region N1 give R(N1) = 1.9 / \sf ~pc. Comparison with
the
observations  indicates that a lower \ff ~should characterize region N1 
because N1 is at a larger distance  from the nucleus.

The modeling of the spectrum observed from region N3  shows that the emission
from gas
photo-ionized and heated by black body radiation corresponding to \Ts = 5 $10^4$ K dominates.
To  estimate the distance r of the nebula from  the bb radiation source 
we  recall that $ N_{ph}$ ($\rm \rho / r)^2$ = U n c,  
where $ N_{ph}$ is the Planck function, $\rho$ the radius of the stars,
r the distance from the nebula, U the ionization parameter, and n the gas
density.  
 A bb temperature of 5 $10^4$ K corresponds to $ N_{ph}$ = 6. $10^{24}$ $\rm
 cm^{-2}\, s^{-1}$
Adopting $\rho$= $10^9 - 10^{10}$ cm, U = 0.001 and n = 20 \cm3 
(Table 1),
the distance of the nebula from the bb source is $\sim$ 0.03 - 0.3 pc .
This distance is very small compared to the dimension of region N3
and shows that the stars are within this region.
The same conclusion applies to other bb radiation dominated regions.

\subsection{The circumnuclear region}

The line spectra corresponding to the observations of  the circumnuclear region
at P.A.= $90^o$ are given in Table 3. 
The spectra are poor in number of lines, but
the lines are the most significant ones. The best fitting model (BB6)
shows the presence of stars with temperatures of $\sim$ 5 $10^4$ K, in 
agreement with
those found in the nuclear region, however, the geometrical
thickness of the clouds is higher ($\geq$ 1 pc), and the density is lower 
(\n0 = 10 \cm3).

The spectra observed at P.A.=$160^{o}$ are given  in Table 4.
The comparison of the observed spectrum from region A1 
with the calculated spectra  shows that the emitting gas
has a low density, \n0 = 20 \cm3 and is ionized by a bb radiation  with \Ts = 7
$10^4$ K  and shocks of 200 \kms (model BB7). 
The spectra observed in the A2 and A3 regions are roughly fitted by models with a 
density higher by a factor of 10 and slightly lower bb temperatures (models 
BB8). Model BB8 is calculated adopting that the bb sources are beyond the 
circumnuclear region (the bb radiation reaches the external edge of the clouds, 
the very shock front edge).
Interestingly, region A4 shows a different situation. In fact, A4 is located
in the very outskirts of the galaxy. Here (see model BB1)
both densities (\n0 = 60 \cm3) and velocities (\Vs = 50 \kms) are low  
because the physical conditions are merging with those of the ISM.
Radiation from the stars, which reaches the internal edge of the
clouds, is very weak but still affecting the spectrum. The clouds are rather 
extended, with D = 0.2 pc.

The spectrum observed from region B is also roughly fitted by model BB8. 
However, the contribution of model BB9 calculated with a higher \Vs  
($\sim$ 300 \kms), a lower \n0 ($\sim$ 30 \cm3), and a bb
temperature of 2 $10^4$ K could improve the fit of some lines.
This model is calculated considering that the radiation from the bb source is
acting on the edge of the clouds opposite to the shock front. 
In other words, these stars are located in a region of the galaxy internal to B.

\subsection{Spectra observed at P.A.=$11.5^o$}

The data from the observations of Shields \& Filippenko at P.A.= $11.5^o$
are given in Tables 5. Observed and calculated lines are given
as ratios to \Hb = 1.
As noticed above, NGC 7130 is an hybrid case of an AGN coexisting with the
starburst.
In Table 5a we show the best fit obtained by pl models, 
representing the AGN,
while in Table 5b bb models are shown, representing the 
starburst.
 Notice that these 
models refer to the narrow line spectrum, which  must be added to the
broad line spectrum for the final result.
 In the other regions where the
deblending is less clear we have used for modeling the summed value of 
[O III] 5007 and [O III] 4959.
The \Hb ~absolute fluxes (\Hb*) are observed at earth, while model results refer
to the nebula. Therefore, a direct comparison is senseless.

 The calculated velocities range between 200 and 280 \kms in agreement with
the observed FWHM of the line profiles. Only in the region observed
at -2.4" the velocities (and the densities) are lower than the observed value
(\Vs $\sim$ 200 \kms), \Vs $\sim$ 100 \kms and \n0 $\sim$ 100 \cm3. 
In the western side of the galaxy
the densities are lower than in the eastern one, out of region at -6.0"
which resides close to the spiral arm.
In the regions located at -3.6" and -4.8" the densities are similar to those
obtained in  the nuclear region N3 and in the circumnuclear regions
A1, A2, B1 (P.A.= $90^o$), and A1 (P.A.=$160^o$).

The fit of the [S II]  line ratios is not always exact. 
As observed by Shields \& Filippenko (1990), the [N II] 6548+6584/[S II]
6716+6730 line ratios in this object are  unusually high ($>$ 2).
 
 Recall that one goal of this work is to give a more complete
picture of the physical conditions in  NGC 7130, considering, particularly, 
the contribution of shock waves.
Therefore, we  focus on the line ratios which  are less easily explained
by pure photo-ionization models, and we
present in Figures 4a and b the results of
[NII]/[SII]  ratios,  calculated by composite models, as function of the 
input parameters corresponding to both bb and pl models, accounting for the shock.
In the diagrams  we compare  model results with the data  of NGC 7130 
observed at P.A.= 11.5$^o$
in order to constrain 
the range of the input parameters. Moreover, the diagrams can be used generally
to explain these line ratios in other objects.

The highest and lowest observed ratios  are shown by solid lines, whereas the 
average value is
shown by a dashed line. Notice that the average value is rather high,
because only in one position  (+6.0", Tables 4) [NII]/[SII]=1.1.
Fig. 4a shows that the  ratios  calculated by bb models are  mostly included within
the observed limits. The   input parameters over the whole region covered by
the slit are constrained  to :
\n0 $\leq$ 200 \cm3,  D $\leq$ 10$^{18}$ cm, 100 \kms $\leq$   \Vs $\leq$ 250 \kms,
and -3 $\leq$ U $\leq$ -2. 
 These results are in agreement with the parameters adopted to fit the data in Table 5b.
Regarding pl models (Fig. 4b), the highest [NII]/[SII]
line ratios can be obtained adopting 9.5$\leq$ log \Fn $\leq$ 10.3 with a few points
also at log \Fn = 11, 10$^{16}$ cm $\leq$ D $\leq$ 10$^{18}$ cm, with  a few points also at D$>$ 10$^{19}$ cm,
\n0 $\leq$ 200 \cm3, and \Vs $\sim$100-200 \kms. 
These values were adopted to fit the models in Table 5a.

The  results can be explained considering that the first ionization
potential of sulfur, IP(S)=10.36 eV, is lower than that of nitrogen,
IP(N)=14.53 eV. The larger is D,
the larger the
 region of low temperature gas inside the clouds,  so [S II] lines are stronger.
On the contrary, the stronger is \Fn (and U), the higher is the temperature of
the gas
in a large zone, so [N II] lines are stronger.
The heating of the gas is also  dependent on the shock velocity, \Vs,
but the  density plays an important role in the cooling rate, so, the trends
of  [N II]/[S II] as functions of \Vs ~and \n0 are less monotonic.
However, these diagrams give only a rough indication because the contribution
of  the broad line  components must be summed up consistently.

The accurate modeling of the spectra  at P.A. = 11.5$^o$ confirms the 
previously found result that  the AGN prevails in the inner regions 
between -2.4" and +2.4", whereas the starburst prevails in the regions 
farther  from the nucleus.
However, the two regions observed at -4.8" and +6.0" could also be fitted by pl
models
calculated with a flux  intensity from the AC lower than that used to fit the
nuclear spectrum. A few lines are observed in  those regions, consequently,
the models are less constrained.
Indeed, the flux intensity decreases with distance from the nucleus, while,
logically, the values of U  assumed in the bb models (Table 5b) 
do not show
an outward decreasing trend, because they depend on the location of the 
starbursts, either in the nuclear region  or in the arms.

The complex  nature (AGN + SB) of NGC 7130  is schematically
illustrated in Fig. 5, where the ionization parameter U is given throughout
the galaxy at P.A. = 11.5$^o$ as a function of distance from the center. 
U is lower in the nuclear region than in the  circumnuclear regions.
This indicates that the ionizing sources  in the nuclear region  and in the circumnuclear region
have a different nature.
Recall that  for AGNs, U $\propto$ \Fn /n, where n is the density
after compression (n/\n0 $\ge$ 10) downstream of the shocks  accompanying the cloud motion.
The photoionizing radiation flux is diluted by  distance from the AC.
 The clouds in the nuclear region are at a distance of $>$ 2 pc from the ionizing source (the  AC),
which is higher by a factor  $\ge$ 10 than  the distance of the clouds from the ionizing source (hot
stars) in the SB region
(\S 3.2). Moreover,  n in the  SB case corresponds to the preshock
density, therefore, a higher U is found in the starburst regions.
Higher \Fn ~in the nuclear region would correspond to  higher U, but  higher \Fn ~are
excluded by the fit of the line spectrum.
Interestingly, the  ionization parameter  in the outskirts of the galaxy  show  maxima at -5"
and +5" indicating the location of the SB in the regions in which the slit crosses the SB ring.

\subsection{The broad line component}

 FWHM observations of the line profiles in the nuclear region
(RRBR, Shields \& Filippenko 1990)
indicate velocities  up to 1000 \kms. The analysis of the line profiles 
shows that the contribution  of the broad component to the 
[O III] line intensity
is  comparable to that of the narrow one (RRBR). 
The measurements of [N II] and [S II]
line profiles by RRBR and Shield \& Filippenko (1990)
show, however, that the broad line contribution is  less important for these lines.

Two high velocity shock (\Vs = 1000 \kms) models, HV1 and HV2, are 
tested. They are characterized by
the different preshock densities (Table 6).
The models were selected among those best fitting both the line ratios and the 
continuum SED simultaneously (cf \S 4). 
High velocities in the NLR of AGN are generally related with high densities
($ \geq 10^{5}$ \cm3) which are required to explain the low  \ROIII
($\equiv$ [O III] 5007/[O III] 4363) line ratio. 
Moreover, a stratification of densities increasing towards the broad line region
 characterizes the narrow line region of AGN.
Model HV1  has \Vs = 1000 \kms and \n0 = 1000 \cm3,
is shock dominated (s.d., \Fn=0) with a high d/g (5 $10^{-12}$).
The high dust-to-gas ratio prevents the flux from the AC to reach the gaseous
cloud,
so that ionization and heating of the gas are due only to the shock.
Compression is high and the downstream density reaches 
$\rm n_{max}$= 2 $10^{5}$ \cm3.
The \Hb ~absolute flux is low because the  rapid cooling downstream reduces 
the  $\rm H^+$ region.
On the other hand, in composite models
which account for the shock, low \ROIII ~are produced by emission from gas 
heated by the shock to high temperatures and very high densities are not
necessary to explain the low \ROIII ~line ratio.
 In the case of relatively low densities, high velocity models may 
represent jets  which are not decelerated by collision with high density
clouds.  Model HV2 in Table 6  characterized by \n0 = 50 \cm3 
represents this case.
In Table 6 the results of the line intensity ratios to \Hb = 1  calculated by
high velocity 
models are  compared with the observations in the inner nuclear region N2. 
Model pl3 
which better explain the narrow line spectrum (Tables 1
and 2) also appears in Table 6.  
The observed line ratios [Ne V]/\Hb ~and [O II]/\Hb ~from
Shields \& Flippenko (1990) are included in the table.

\bigskip

In summary, both a power-law radiation from the AC and bb radiation from
stars are confirmed as ionization sources in NGC 7130. 
 Shocks of $\sim$ 200 \kms are always present. Power--law radiation dominates
in the inner 2 arcsec; black-body radiation prevails in the outer regions.
The broad line component detected in the nucleus suggests the presence of
a high-velocity, shock dominated component, either with high densities indicative
of stratification as we approach the broad line region or low densities 
related to jets.

Notice that whenever different regions of emitting gas coexist in one object
the observed spectra contain the contribution of all of them. The most
probable model must consistently account for all lines and continua.

\section{The spectral energy distribution}

The purpose of this section is to check whether the continuum produced by the
previously derived models is consistent with the observed spectral energy
distribution (SED). The continuum SED emitted from the NLR of a group of 
Seyfert 2 galaxies (Heckman et al. 1995) was investigated by 
Contini \& Viegas (2000). We adopt their point of view
and  assume that the continuum SED is essentially radiation 
reprocessed by the  clouds: bremsstrahlung from gas in the UV-optical 
range and reradiation  by dust  in the infrared range (Viegas \& Contini 1994).
Alternatively, the optical-UV SED of the continuum can be  explained by the 
distribution of black body radiation from relatively high temperature stars.
This case, however,  seems less reliable, as discussed previously by 
Contini \& Viegas (2000, \S 2.2.2.3).

The SED data represent the integrated emission from 
the whole galaxy observed at each frequency. 
The UV to optical continuum data are taken from  
Kinney et al. (1993), De Vaucouleur et al (1991),
Lauberts \& Valentijn (1989); far-infrared data are taken from 
Moshir et al (1990).
The datum at 1.4 GHz  comes from Whittle et al. (1992) and the other data in the
radio range from Bransford et al (1998). The upper limit
at 1 keV is given by Rush et al. (1996).


We compare in Figs. 6, 7, and 8 the observed continuum SED  with single-cloud models which
explain the observed line spectra (\S 3).
We showed in the previous section that three components are needed in order
to model the emission line ratios: i) clouds ionized by  power-law radiation
from the AC, ii) clouds ionized by bb radiation from stars and iii) high
velocity clouds. All these components may contribute to the observed continuum.
We first analyse the contribution to the SED given  by the single-cloud
models discussed above, for each component separately. We then show
that the observed SED is consistent with a weighted sum of these single-cloud
SEDs. Given the few points available from observations and the heterogeneous
way in which they were obtained, this should be considered as a
consistency check rather than an exact fit.

The three radiation components (synchrotron emission
in the radio range, dust emission in the IR, and bremsstrahlung
emission from the gas) are shown separately in the figures.
No data are available in the UV range and we assume that radiation from the AC
is completely absorbed.

The data are measured at Earth, whereas the models calculate the flux at
the nebula.
To exactly fit the data, the  intensity of the calculated continuum flux is 
multiplied by a  scaling factor  which accounts for the number of the typical 
emitting clouds,  
for the filling factor  and, above all, for the square R/d ratio, 
where R is the distance of the emitting region from the center of the galaxy and 
d is the distance of the galaxy from Earth. 

Notice that emission by dust is represented by the peaks with maxima  at 
about 2 $10^{12}$ Hz.
The peak height constrains the d/g value, after normalizing the bremsstrahlung 
from the gas, while the frequency corresponding to the maximum peak depends on
the temperature of the grains.
Gas and dust are coupled within the clouds.
The grains are heated by radiation and by
collisions (see Viegas \& Contini 1994).
In the presence of shocks the gas reaches the maximum 
temperature  in the immediate
postshock region before rapidly cooling downstream, therefore, the maximum
temperature of the
grains is determined by the shock velocity. 
In other words, reradiation by dust
from models with the same \Vs ~peaks at about the same frequency. 
Consequently, the datum in the mid-IR, at $\nu = 1.2 \, 10^{13}$ Hz, 
will be fitted  by models characterized by a higher \Vs.

\subsection{ Models calculated by  a power-law radiation}


The SED of the observed continuum is compared with 
models PL1, PL2, and PL3 in Fig. 6.
Fig. 6 shows that model PL1 (dashed lines) and model PL2 (solid lines) better 
reproduce the data in the optical-near UV range. 
Model PL3  (dotted lines) reproduces roughly the optical-IR data but 
underestimates the data at higher frequencies in the near UV.

The  flux  from the AC  seen throughout the clouds is mainly 
accounted for by model PL3 
(dotted line in  Fig. 6 at the highest frequencies), so the other 
models are not represented in the figure.
All the models overestimate the data in the radio range and show a different
trend  (Figs. 6 and 7), indicating that 
bremsstrahlung is partly absorbed at these  wavelengths (Osterbrock 1989). 
On the other hand, the data are well fitted by synchrotron radiation
created at the shock front by Fermi mechanism,
in agreement with Bransford et al. (1998) who
attribute the radio emission to either
shocks or star formation within the bar at P.A.= 0$^o$.

\subsection{Models calculated by black body radiation}

 
 Models (BB2, BB3, BB4, BB5, BB6, BB7,
BB8, and BB9) are compared with the data in Fig. 7.
Model BB9 is represented by the dashed lines and BB8 by the solid 
lines, all the other models by dotted lines. 
The models are normalized to the highest frequency data in the near UV.

The trend of  bremsstrahlung  in the optical-near UV region (10$^{14}$- 10$^{15}$ Hz)
is different from the observed one. However,   model BB9
explains the upper limit in the soft X-ray range.
Model BB8 shows a maximum peak of the bremsstrahlung emission at
about 1.6 $10^{14}$ Hz. 
The peak in the soft X-ray region of model BB9 is at higher frequencies compared 
with the other models because \Vs ~is higher, and the mid IR datum at 
1.2  $10^{13}$ Hz is also well reproduced.  
The fit of the far IR  and mid IR data settle a d/g of about 1-2 $10^{-14}$.

The data between 10$^{13}$ and 3 10$^{14}$ Hz could be reproduced 
by black body radiation from stars at a temperature of 3000 K which 
represents the contribution from the old star population background
as was noticed in other Seyfert 2 galaxies (Contini \& Viegas 2000).

\subsection{High velocity models}


 High velocity shocks  are generally
invoked to fit the data in the soft X-ray domain and reradiation by dust in the
mid IR.
Figs. 6 and 7 show that the  upper limit in the soft X-ray  range
is roughly fitted by radiation from the AC and/or bremsstrahlung from a 300 \kms
velocity shock (model BB9). The two
data in the near IR  can be reproduced by reradiation of dust (at 1.2 $10^{13}$ Hz,
model BB9)
and by bremsstrahlung of gas reached by a strong power-law radiation 
flux (at 2.5 $10^{13}$ Hz, model PL3), respectively. The simultaneous fit 
of data in both the soft X-ray and IR frequency domains, however, does not 
exclude emission from hot gas and dust
corresponding to  high velocity shocks.
The continuum SED of models HV1 and HV2 appears in Fig. 8.
The two models peak at slightly different frequencies owing to
the strong effect of the density. The  high density of model HV1 enhances
cooling thus reducing the emission from hot gas.

\bigskip

Summarizing, the comparison of Figs. 6, 7, and 8 leads to the following results.  

The upper limit in the soft X-ray is 
ambiguously reproduced by the
flux from the AC (Fig. 6), and bremsstrahlung from gas heated by shocks of \Vs $\geq$ 300
\kms (Figs. 7 and 8). 

Models dominated by pl radiation, which represent the AGN 
explain the continuum dataset in the optical -  near UV range, while
most of the models dominated by bb radiation, which
represent the starburst,  show a trend different than that of  the observed 
continuum  in the  frequency range between 10$^{14}$  and 10$^{15}$ Hz. 

The data in the optical range
are better explained by adding   the contribution of the background old stellar
population
(\Ts = 3000 K) to the bremsstrahlung from bb dominated models (Fig. 7).

Emission in the IR is explained by reradiation of dust which is mainly
collisionally heated by the shocks. 

In the radio range  self absorption  reduces  bremsstrahlung radiation  at
long wavelengths   and the data are  explained by
synchrotron radiation created by Fermi mechanism at the shock front.

\subsection{The multi-cloud spectrum}

It was shown in the previous sections that the spectra emitted  from the
different regions of the NGC 7130 galaxy cannot always be modeled
by a single-cloud spectrum. Therefore, we compare in Table 6
 the spectra
which result from the weighted sum of single-cloud spectra to those observed
in the nuclear region N2. The continuum observed from
the NGC 7130 galaxy is compared with a multi-cloud model in Fig. 9.

In the top of Table 6 the line ratios to \Hb\ are given; in the 
middle
of the table the input parameters appear, followed in the bottom by
the weights (W) adopted in the average sum and the relative
contribution of single models  to the [OIII] 5007 line ($\rm P_{[OIII]}$).

We assume that the narrow line contribution to the  spectrum emitted from the N2 region is 
represented by model PL3 because  bb dominated spectra are relatively weak.
The [Ne V]/\Hb ~and [O II]/\Hb ~ observed ratios come from 
Shields \& Filippenko (1990).  Gonz\'alez Delgado et al give [Ne V]/\Hb = 0.6.

 The weights of the broad and
narrow components which give the best fit of the weighted
sum  (AV1) to the N2 spectrum are given in Table 6, 
W(PL3) : W(HV1) :: 1 : 10.
The fit of the line spectrum is within the observational uncertainties.
A slightly higher weight for HV1 should improve definitively the fit to the
observed [O III] 4363 line. 

To respect the upper limit in the soft X-ray range,
the bremsstrahlung emission intensity of model HV1 is reduced by four orders of
magnitude relative to model PL3 in Fig. 9,  whereas the best average of the line 
spectrum is obtained by raising the HV1 flux by a factor of 10  
relative to PL3 in Table 6. 
Two different weights in the line and continuum averages are not contradictory
considering
the strong absorption of bremsstrahlung radiation in the X-ray domain.
 Assuming a cross section, $\sigma_{abs}$ = 2 $10^{-22} \, \rm  cm^2$ (Zombeck
 1990) at about $10^{17}$ Hz, the optical depth 
$\tau$ = $\rm n_{max}$ D $\sigma$ \ff  results 120 \ff for model HV1.
 A  reduction  by a factor of $10^{-5}$  of the bremsstrahlung emission in the
 X-ray domain  gives \ff $\sim $0.04.
Dust is heated at high temperatures by collision with  gas.
Reemission by dust in the near IR is reduced by the same
factor as in  the X-ray, because there is mutual heating between dust and gas.

As discussed previously, an alternative model for the broad line component is 
characterized by a low preshock
density, \n0 = 50 \cm3.  The line spectrum calculated by model HV2 is given in
Table 6 and the 
calculated continuum SED is presented in Fig. 8. 
 The averaged sum of PL3  and
HV2 (AV2)
which gives the best fit to the line spectrum  implies comparable weights,
namely,
W(PL3) : W(HV2) :: 1 : 1 (Table 6). 
The [O III] 4363/ \Hb ~line ratio is underestimated,
therefore,  the  averaged model 
AV2 is less valid than the averaged model AV1.

The final fit to the continuum SED by a multi-cloud averaged model
is presented in Fig. 9. The continua of single-cloud models are summed up
relatively to their weights. 
Model PL2 representing the nuclear region N1 is added to the model 
 PL3 which represents the nuclear region N2 with a weight lower
by a factor of 6. This indicates a larger distance from the
AC. Owing to its lower weight, model PL2 is not included
in Table 6.
The weights of bb radiation dominated models which
represent the circumnuclear starburst are by a factor  $<$ 10 lower than the
weight
of model PL3 which prevails in the nuclear region and represents the AGN.
Notice, however, that  the contribution of
clouds photo-ionized by bb  radiation to the  continuum spectrum cannot be neglected 
because the line spectra calculated by bb models explain 
most of the observed line spectra, particularly, in the circumnuclear region.
They  are, therefore, included in Fig. 9.

The multi-cloud model presented in Fig. 9 shows
a mean ultraviolet-optical slope between a rest wavelength 
of 1910 A and 4250 A of $\gamma$ $ \sim $ 1
($\rm F_{\lambda} \propto \lambda^{\gamma}$)
and a power-law index in the power-law continuum between 1230 A and 2600 A,
 $\beta$ = -0.54,  in agreement with $\gamma$ = 0.9 
and $\beta$ = -0.5, respectively, given by Heckman et al. (1995, Table 1).
Our analysis therefore indicates  that the ultraviolet-optical continuum
is dominated by  radiation from the AGN. It should be however noted that 
Storchi-Bergmann et al. (2000) concluded, from the comparison with
spectral models, that the 3500-4100 A spectrum observed in a 2"x2" aperture
is consistent with the combination of a bulge template with light from young
and intermediate-aged stars. 

\section{Discussion}

In previous sections  the distribution of the densities
and of velocities throughout the galaxy are calculated. 
The distribution of the densities
can be related to the location of the SB, whereas the distribution of velocities
is  consistent with the velocity field in the NLR of AGN.

The preshock densities calculated in each region of NGC 7130 are
marked in Fig. 10. The highest densities,
which are not displayed in Fig. 10 for sake of clarity, are 
found in the nuclear region and reach 1000 \cm3.
These densities are typical of the high velocity clouds in the inner
narrow line region of the AGN. Densities (preshock) of about 200 \cm3 appear
in the nuclear N2 region and in the circumnuclear spiral arms.
 In the region between them the densities
range from  10  to 50 \cm3.

More particularly,  in the
circumnuclear region at P.A.= $160^o$, 
 the preshock densities calculated in regions A1 and A4
(see Table 4) 
are lower by a factor of about 10  than those  in regions A2 and A3
which contain the starbursts and correspond to the spiral arm.
The calculated ionization parameter U for A1 and A4 is lower
than for A2 and A3 by a factor  $\geq$ 10, indicating that the clouds
in these two regions are farther from the ionizing source (\S 3.2),
 i.e. the starbursts, which in fact are located only into regions A2 and A3.
The same is valid for regions A1, A2, and B1 observed at P.A. = $90^o$. 

 The  velocity distribution throughout NGC 7130 shows
  relatively high velocities (1000 \kms) close to the
active center, a large region (nuclear and circumnuclear) dominated
by bulk velocities of about 200 \kms and a
 velocity definitively lower in  region A4 located 
in the very outer edge of the galaxy.
This is in agreement with  the large scale  radial motions generally
present in the narrow line region of AGN.

Notice that  compression
is strong (up to a factor $\geq$ 10) in the cloud downstream region, 
depending on \Vs ~and \n0, triggering star formation.
Star formation presumably  occurs once the density crosses the threshold 
critical value. As observed by Kennicutt (1989)
the nonlinear increase in the star formation rate near the threshold may
provide a straightforward explanation for the strong concentration of star 
formation in the spiral arms of many galaxies.
NGC 7130 shows strong star formation in the nuclear region and in the 
spiral arm, as expected from the  ratio of the density in the central region 
and in the arms to that in the interarm regions. The ratio
is larger than 4, and thus consistent with a Schmidt law for normal 
values of the power-law index ($n \sim$2).

\section{Concluding remarks}

We present new consistent calculations of both the line and continuum spectra 
of the composite galaxy NGC 7130. The analysis throughout the galaxy shows
 that  the nuclear region is dominated by
spectra emitted from gas  ionized by a  power-law radiation  from
the AC. The effect of the shock is also important
and is revealed  from both the SED of the continuum and the line spectra.

Shock velocities of about 200 \kms are found in agreement with the FWHM of the
narrow component observed in the line profiles. 
The emission from high shock velocity  clouds (\Vs = 1000 \kms), 
improves the agreement of calculated to observed  continuum and line spectra,
 particularly, the fit of the [O III]4363/ \Hb
~and [NeV]3426/\Hb ~line ratios, which are underpredicted by pure photoionization
models.
The high velocity clouds are  close to the AC and are characterized by 
a high dust-to gas ratio ($>$ 10$^{-12}$), while
the dust-to-gas ratio is about $10^{-14}$ throughout the galaxy.
The high d/g in the central region of the galaxy 
    prevents  radiation  from reaching the high velocity, high density
gas which emits the broad  component of the lines.

Massive stars with temperature of 5-7 $10^4$ K  photoionize and heat the gas 
in the circumnuclear regions.
An old star population (\Ts = 3000 K) background contributes  to the SED
of the continuum in the optical - near IR domain.
The SED of the continuum in the optical range  is dominated by free-free
emission from gas reached by a power-law radiation flux from the AC,
while bremsstrahlung from clouds ionized and heated by thermal radiation
from young massive stars contributes in the near-UV range.
The stars in the nuclear region are close to the emitting  clouds at a distance
of $\sim$ 0.1 pc.

Finally, the agreement of results between the two codes CLOUDY, which accounts for pure
photoionization (adopted by RRBR), and SUMA, which accounts for photoionization and shocks,
 concerns only the intermediate and low ionization lines,  regarding the values of
the spectral index ($\alpha$ = -1.5  adopted  by both codes),  the ionization parameters, 
(log U $<$ -3  adopted by SUMA  and -3.5 by CLOUDY in the nuclear region and
log U = -3 adopted by both codes in the circumnuclear region), the type of radiation
pl and bb, prevailing in the nuclear and circumnuclear regions, respectively,
and the N/H relative abundance ($<$ 1.5 cosmic by SUMA and  twice cosmic by
CLOUDY).
The fit of the average model AVC (Table 6)  calculated by CLOUDY to the
 observed spectrum, is  excellent but implies a very high 
contribution of the high density component (model CL1) to the [O III] lines. 
If the high density gas represents the broad line component
this  gives a contribution of 90 \% to the [O III] 5007 line and 96 \% to the [O
III] 4363 line, 
making the narrow contribution (model CL2) to the line profile actually unobservable.
In the average   models
calculated by SUMA the contributions of the broad line component to
[O III] 5007 and [O III] 4363 are  $<$ 13 \% and 72  \%, respectively.

Moreover, the fit to [NeV] 3426/\Hb ~line ratio is very poor in the model
calculated by CLOUDY.  The [NeV] line  corresponds to a relatively high ionization 
level,  indicating that gas at a relatively high temperature
should be present, which justifies gas heating by shocks.

\bigskip

Acknowledgements

 We are grateful to the referee for very constructive criticism. 
M.C. is grateful to the Astrophysikalisches Institut of Potsdam
for warm hospitality. We acknowledge the access to the HST Science Data 
Archive and the ESO/ST-ECF Archive Facility for the retrieval of the HST 
WFPC2 image and the GNA of CNR for financial support.  

\newpage

{\bf References}

\bigskip

\vsize=26 true cm
\hsize=16 true cm
\baselineskip=18 pt
%
\def\ref {\par \noindent \parshape=6 0cm 12.5cm 
0.5cm 12.5cm 0.5cm 12.5cm 0.5cm 12.5cm 0.5cm 12.5cm 0.5cm 12.5cm}
\ref Allen, C.W., 1973 in "Astrophysical Quantities" (London: Athlone) 
\ref Baldwin, J.A., Phillips, M.M., \& Terlevich, R. 1981 PASP, 93, 5 
\ref Best,P.N., Longair, M.S., \& Rottgering, H.J.A. 1997 MNRAS, 286, 785 
\ref Bransford, M. A. et al. 1998 ApJ 497, 133
\ref Burbridge, G.R., Gould, R.G., \& Pottash, S.R.   1963, ApJ, 138, 945 
\ref Contini, M. \& Viegas, S. M. 2000, ApJ, 535, 721
\ref Contini, M. \& Viegas, S. M. 2001, ApJS, 132, 211
\ref Contini, M. Prieto, M.A. \& Viegas, S. M. 1998 ApJ 505, 621
\ref De Vaucouleurs, G., De vaucouleurs, A., Corwin Jr, H.G. et al.
1991 Third Reference Catalogue of Bright Galaxies, Version 3.9
\ref Ferguson, J.W., Korista, K.T., \& Ferland, G.J. 1997, ApJS 110, 287 
\ref Ferland, G.J. 1996, Hazy, a brief introduction to Cloudy, University of
 Kentucky, Department of Physics and Astronomy Internal Report
\ref Gonz\'alez Delgado, R.M. et al. 1998 ApJ, 505, 174
\ref Heckman, T., Krolik, J., Meurer, G. et al. 1995 ApJ, 452, 549 
\ref Heckman et al. 1995, ApJ, 452, 549
\ref Hutchings, J.B. et al. 1999, AJ, 118, 210
\ref Kennicutt, R.C. 1989, ApJ, 344, 685
\ref Kinney, A.L., Bohlin, R.C., Calzetti, D. et al 1993 ApJS 86,5
\ref Lauberts, A. \& Valentijn, E.A. 
The Surface Photometry Catalogue of the ESO-Uppsala
Galaxies, 1989, Garching bei Munchen, ESO
\ref Levenson, N.A., Weaver, K.A., \& Heckman, T.M. 2001, ApJS, 133, 269
\ref Moshir, M.,Koplan, G.,Conrow, T. et al. 1990
Infrared Astronomical Satellite Catalogs, 1990,
The Faint Source Catalogue, Version 2.0
\ref Osterbrock, D. E. 1965, ApJ, 142, 1423
\ref Osterbrock, D. E. 1989, "Astrophysics of gaseous Nebulae
and Active Galactic Nuclei" (Mill Valley: University Science Books) 
\ref Phillips, M.M., Charles, P.A., \& Baldwin, J.A. 1983 ApJ 266, 485
\ref Radovich, M., Rafanelli, P., Birkle, K., \& Richter, G. 1997 Astron. Nachr., 318, 229
\ref Rush, B., Malkan, M.A., Fink, H.H., \& Voges, W. 1996 ApJ, 471, 190
\ref Shields, J.C. \& Filippenko, A.V. 1990 AJ, 100, 1034
\ref Storchi-Bergmann Th., Raimann D., Bica E.L.D. \& Fraquelli H.A. 2000 ApJ,
544, 747
\ref Viegas-Aldrovandi, S. M. \& Contini, M. 1989 ApJ 339, 689
\ref Viegas, S. M. \& Contini, M. 1994 ApJ 428, 113
\ref Whittle, M. 1992 ApJS 79, 59
\ref Zombeck, M. V. 1990 in "Handbook of Space Astronomy and Astrophysics"
Cambridge University Press, p.199

\newpage

{\bf Figure Captions}

\bigskip

Fig.1

Positions along the slit 
 overplotted on a HST WFPC2 image of NGC 7130 (F606W filter).

\bigskip

Fig. 2

  The figure display the two models here adopted: 
  (a) the inner edge of a cloud is ionized by radiation from the active 
  nucleus and/or from hot stars, the shock is acting on the outer edge; 
  (b) the radiation from stars and the shock front act on the same edge of the
  cloud.

\bigskip

Fig. 3

Results of inflow models are compared with the data (empty triangles).
Filled triangles : \n0= 50-100 \cm3, log \Fn = 10 - 11,
D=10$^{18-19}$ cm.
Filled  squares : the same as for filled triangles with \n0=200 \cm3.

\bigskip

Fig. 4

The diagrams represent the calculated [N II]/[SII] ratios as a function of the
main input parameters, U, \Fn (in $\rm photons \, cm^{-2} \, s^{-1} \, 
eV^{-1}$), D (in cm), \n0 (in \cm3),
 and \Vs (in \kms), for models which better fit
the observational data. Filled squares refer to models calculated in the case
that the radiation flux and the shock act on opposite edges of the cloud.
Open squares refer to radiation and shocks acting on the same edge.
a) : models calculated by  black body radiation from the starburst.
b) : models calculated by  power-law radiation from the AC.
Solid lines indicate the highest and lowest [NII]/[SII] oserved line ratios
from Shields \& Filippenko (1990). The dotted line indicate the average
value.

\bigskip

Fig. 5

The distribution of the ionization parameter U throughout the
galaxy at P.A. = 11.5$^o$. Filled circles : pl models;
filled squares : bb models.

\bigskip

Fig. 6

 The comparison of model calculations with the observations 
of the SED of the continuum (filled squares) for
 power-law radiation dominated models.
PL1 is represented by short dashed lines, PL2 by  solid lines, and
PL3 by dots.

\bigskip

Fig. 7

The same as Fig. 6 for
 black body radiation dominated models. BB8 is represented by solid lines,
BB9 by dashed lines, and all the other models by dots.
The dash-dotted line indicated black body radiation from old stars
(see text).

\bigskip

Fig. 8

 The comparison of  high velocity model  SED with observations
(filled squares). 

\bigskip

Fig. 9

 The best fit of model weighted sum (solid lines) to the observed data
(filled squares). The dashed line shows that absorption is present in the low
frequency range. Dotted lines represent single models (PL2, PL3, BB5, 
BB6, BB7, BB8, BB9 and HV1).

\bigskip

Fig. 10

Positions along the slit and preshock density calculated by the models
 for the regions observed by RRBR and by Shields \& Filippenko 
are overplotted on a HST WFPC2 image of NGC 7130 (F606W filter).

\newpage

\begin{table}
\centerline{Table 1}
\centerline{The spectra in the nuclear region (P.A.=$90^o$)}
\small{
\begin{tabular}{l l llll ll ll} 
\hline
\hline
 Line & N1 & PL1  & PL2 &  BB2 & N2 & PL3 & BB3 & N3& BB4     \\
$r_c$(")       &  -2.0   &- &- &- &  0.1   &- &- &  2.4      &-\\
$\Delta r$(")   &   2.1   &- &- &- &  2.1   &- &- &  2.4      &-\\ \hline\\


\ [Ne III] 3869+ & 1.35 & 1.23 &0.8&1.0&1.07 & 0.9&1.47& - &0.6      \\
\ [S II] 4073+   & - & 0.17 &0.08&0.03&0.28&0.22 &0.05& - &0.02       \\
\ [O III] 4363 & 0.03 & 0.04&0.03&0.4 &0.13&0.05&0.8& - & 0.3    \\
\ He II 4686 & 0.25 & 0.17 &0.13& 0.04& 0.15 & 0.04 &0.04& - & 0.02   \\    
\ [O III] 5007+& 6.9  & 6.70 &6.0 & 6.23& 8.3 & 9.24 &10.0& 4.4 & 4.5    \\
\ [N I] 5200+& 0.30 &0.23 &0.10&0.02& 0.12& 0.10 &0.01& - & 0.01     \\         
     
\ He I 5876 & - & 0.13 &0.13 & 0.12&0.06 & 0.13&0.11& - & 0.13     \\           
      
\ [O I] 6300+ & 0.24 & 1.60 &0.5& 0.04& 0.47 & 0.43  &0.04& 0.21 & 0.10   \\    
        
\ [N II] 6548+ & 3.53 & 4.75 &3.2& 3.0&4.12 & 2.6 & 2.6&2.37&2.0   \\           
      
\ [S II] 6716 & 0.42 & 1.0 &1.0&0.43& 0.4 & 0.7 & 0.4&0.65 &0.35     \\         
      
\ [S II] 6730 & 0.35   & 1.26& 0.9& 0.37 &0.44 & 1.0 &0.37& $\uparrow$& 0.27   
\\   
\ [Ar III] 7136 & - & 0.27 & 0.18 &0.13  &0.07& 0.13&0.15&- & 0.13      \\      
          
\ [O II] 7325 & 0.29 & 0.23&0.11&0.46&0.14 & 0.08 &1.5& - & 0.46   \\           
     
\ $\rm H\beta$ (\erg) &-& 0.017&0.011&1.(-4)&-&0.7&1.3(-4)&-& 1.3(-4)    \\ 
 & & & & & & &&&\\
 \Vs  ($\rm km\, s^{-1}$) & -& 240 &200&200&-& 200&200&-&200         \\
 \n0  ($\rm cm^{-3}$) &- & 100&50& 20&-&200 &50&-&20         \\
 \Ts  ($\rm K$) & -& - &-&7(4)&-&-&3.5(4)&-&5(4)       \\
  U   &- & $<$1(-3) &$<$1(-3)&1(-3)&-&$<$1(-3)&1(-3)&-&1(-3)     \\
  log \Fn   &- &9.7 &9.3&-&-& 11 &-&-&-       \\
  D (cm)&-& 3(17)&6(17)&2.3(17)&-&2.3(18)&3.9(16)&-&4.6(16) \\
  d/g  &- & 5(-15) &5(-15)&6(-17)&-&1(-14) &3(-14)&-&1(-14)  \\
\hline
\hline
\end{tabular}}
\end{table}

\begin{table}
\centerline{Table 2}
\centerline{The spectra in the nuclear region  (P.A.=$160^o$)}
\small{
\begin{tabular}{l l lllll }     \\ 
\hline
\hline
 Line & N1 & BB5    &N2& PL3 & N3& BB5   \\
$r_c$(")   & -2.8 &-& 0.1 &-& 3.7 &- \\
$\Delta r$(")  &  3.1 &-& 2.6 &-& 4.5 &- \\ \hline\\

\ [Ne III] 3869+ & 4.51 & 4.6 & 1.61 &  1.0& 7.47 & 4.6 \\                
\ [S II] 4073+   & - & -    &0.49&0.22 &-   & -        \\
\ [O III] 4363 & - & - & 0.14 & 0.05 & - & - \\                         
\ He II 4686 & - & - & 0.17 & 0.04 & - & - \\                                   
  
\ [O III] 5007+& 8.77 & 9.60 &8.15& 9.4 & 9.1 & 9.60 \\                
\ [N I] 5200+& - & - & 0.16 & 0.10 & - & - \\                                   
    
\ He I 5876 & - & - & 0.07 & 0.13 & - & -   \\                                 
\ [O I] 6300+ & 0.28 & 0.26 &0.42& 0.43  & 0.24 &  0.26  \\                 
\ [N II] 6548+ & 4.32 & 3.6  &3.86& 2.6 & 2.72 & 3.6 \\                         
     
\ [S II] 6716 & 0.87 & 0.7 & 0.37 & 0.7 & 0.7 & 0.7   \\
\ [S II] 6730 & - & - & 0.4 & 1.1 & - &   - \\
\ [Ar III] 7136 &- & - & 0.06 & 0.09 & - & - \\
\ [O II] 7325 & -& -&  0.13 & 0.08 & - & - \\
\ $\rm H\beta$ ($\rm erg\, cm^{-2} s^{-1}$) &-&1.5(-4) & - & 0.7 & - &  1.5(-4)
\\
 & & & & & & \\
 \Vs      ($\rm km\, s^{-1}$) & -& 200 &- & 200 & - & 200 \\
 \n0      ($\rm cm^{-3}$) &- & 50 & - & 200 & - &  50 \\
 \Ts      ($\rm K$)& -& 7.5(4)&-&-&-&  7.5(4)     \\
  U         &- &5(-4) & - &$<$1(-3) & -&  5(-4) \\
  log \Fn  &- &- &- & 11 & - & - \\
  D (cm)&-& 3(16)&- & 2.3(18) & - & 3(16)\\
  d/g  & - & 1(-14) & - & 1(-14) & - & 1(-14) \\ 
  Ne/H & - & 3(-4) & - & 9.3(-5) &- & 3(-4) \\
\hline
\hline
\end{tabular}}
\end{table}

\begin{table}
\centerline{Table 3}
\centerline{The spectra in the circumnuclear region (P.A.=$90^o$)}
\small{
\begin{tabular}{l l llll l}
\hline
\hline
 Line & A1 & BB6  &A2 & BB6  & B1& BB6     \\
$r_c$(")  & -4.6   &-& -7.9  &-& 8.1 & - \\
$\Delta r$(")   &  3.1   &-&  3.4  &-&  9.2  & - \\ \hline\\

\ [O III]  5007 & 1.24 & 1.8  &- & -    & 1.47&  1.8 \\    
\ [O I] 6300  & -  & - & - & - & 0.2 & 0.12 \\
\ [N II] 6548+ & 1.86 & 1.93 & 1.33 & 1.93 & 2.1 & 1.93 \\
\ [S II] 6716+& 0.54 & 0.70 & 0.47 & 0.70 & 0.58 & 0.70   \\
\ $\rm H\beta$ ($\rm erg\, cm^{-2} s^{-1}$) &-& 1(-4) & - & 1(-4) & - & 1(-4) \\
 & & & & & & \\
 \Vs     ($\rm km\, s^{-1}$) & -& 200 &- & 200 &- & 200   \\
 \n0     ($\rm cm^{-3}$) &- & 10&- & 10 & - & 10 \\
 \Ts      ($\rm K$)& -& 5(4) & - & 5(4) & - & 5(4) \\
  U   &- & 1(-3) &- &1(-3)&- & 1(-3)   \\
  D (cm)&-& 5(18)&- & 5(18) & -&5(18) \\ 
 d/g & - & 1(-14)& - & 1(-14) &- & 1(-14) \\ 
\hline
\hline
\end{tabular}}
\end{table}

\begin{table}
\centerline{Table 4}
\centerline{The spectra in the circumnuclear region (P.A.=$160^o$)}
\small{
\begin{tabular}{l l llll ll llll}     \\ 
\hline
\hline
 Line & A1 & BB7   &A2 & BB8& A3& BB8  & A4 & BB1 & B & BB8  & BB9 \\
$r_c$(")    & -6.8 &-& -11.7 &-& -15.3 &-& -23.2 &-& 10.5 &-&-\\
$\Delta r$(") &  5.0 &-&   4.7 &-&   2.6 &-&  13.1 &-&  9.2 &-&-\\ \hline\\

\ [O III]  5007 & 2.12 & 2.4 & 0.14 &0.20 & 0.14 & 0.20 & 0.07 &0.06 & 0.20
&0.20&0.37 \\
\ [O I]  6300  & -  & - &0.07  &0.02  &0.04&0.02 &- & - &0.04 & 0.02 & 0.19   
\\
\ [N II] 6548+ & 3.86 & 3.30 & 1.33 & 1.0  & 1.1 & 1.0  & 1.27 & 1.6  & 1.49 &
1.0  & 1.33 \\
\ [S II] 6716+& 0.90 & 0.80& 0.57 & 0.20& 0.58 & 0.20 & 0.81 & 1.20& 0.53 & 0.2
& 0.4 \\
\ $\rm H\beta$ ($\rm erg\, cm^{-2} s^{-1}$) &-& 2(-4) & - & 0.05  & - & 0.05 &
-&1.9(-4)&-& 0.05  & 4(-3) \\
 & & & & & & &&&&&\\
 \Vs   ($\rm km\, s^{-1}$) & -& 200 &- & 200 &- & 200 & - & 50  & - & 200 & 300 
  \\
 \n0   ($\rm cm^{-3}$) &- & 20&- & 200 & - & 200 &-& 60 &-& 200& 30 \\
 \Ts   ($\rm K$)& -& 7(4) & - & 5(4) & - & 5(4)&-& 5(4) &-& 5(4) & 2(4) \\
  U  & - & 1(-3) &- &1(-2)&- & 1(-2)&-& 2(-5) &-& 1(-2)& 1(-2)   \\
  D (cm)&-& 4(17)&- & 5(16) & -&5(16) & - & 6(17) &-& 5(16) & 5(17) \\ 
  d/g & - & 1(-14) & - & 1(-14) & -& 1(-14) & - & 1(-14) & - & 1(-14) & 2(-14)
  \\
\hline 
\hline
\end{tabular}}
\end{table}

\oddsidemargin 0.01cm
\evensidemargin 0.01cm

\begin{table}
\centerline{Table 5a}
\centerline{The spectra at P.A.=11.5$^o$ from Shield \& Filippenko (1990)}
\centerline{compared with power-law models (pl)}

\tiny{

\begin{tabular}{l l llll ll lllllllllllllll}     \\ \hline

 Line  &-6"& pl6 &-4.8" & pl5 &-3.6" & pl4 &-2.4" & pl3 &-1.2"& pl2&0" (Nuc)&
pl0& +1.2" & pl2 & +2.4" &pl3 & +3.6" &pl4 &+4.8" &pl5 &+6" & pl6 \\

\hline\\
&& & &  & &  && && && && && && && & \\
\[[NeV]3426&-&- &-&- &-&- &-&- &-&- &0.78&0.04 &-&- &-&- &-&- &-&- &-& - \\
\[[OII]3727& 2.7:&-&1.7&1.2&3.7&-&8.0&7.&3.1&3.1&1.7&1.1&1.0&2.4&3.0&2.4&1.7&-&1.2&-&0.66&1.2  \\
\[[NeIII]3869+ &-&- &-&- &-&- &2.7 &1.3&1.3&1.0 &1.1&1.0 &1.5&0.7 &1.9&0.6 &-&- &-&- &-&- \\                
\[[SII]4071+   &-&- &-&- &-&- &0.96&0.22 &0.31&0.2 &0.26&0.22 &0.43&0.24 &0.93&0.22 &-&- &-&- &-&- \\
\[[OIII]4363 &-&- &-&- &-&- &-&- &-&- &.082&.05 &-&- &-&- &-&- &-&- &-& - \\                         
\[HeII 4686 & - &- &-&- &-&- &0.23&0.13 &0.20&0.12 &0.16&0.04&0.14&.07&0.13& 0.05&.097&- &-&- &-&- \\           
\[[OIII]5007 &0.37&-&0.74&0.75&2.4&-&5.2&5.7 &6.1&7.3&6.0&7.0&5.1&5.3&3.0&3.&1.1&- &0.45&-&0.11&0.2 \\
\[[NI]5200+&-&-&-&-&0.10:&-&0.10&0.&0.16&0.04&0.14&0.1&0.12&0.28&0.13&.007&0.075&-&0.011&-&-&- \\            
\[HeI 5876 &- &-&-&-&0.12:&-&.094&0.11&.042&0.12&.033&0.13&.047&0.36&.044&.09&.078&-&.097&-&.053 & .084\\   
\[[OI]6300+ & .057!&-&.064!&0.13&0.14&-&0.23&0.23&0.26&0.38&.25&0.4&.25&1.26&0.21&0.1&.13&-&.064!&-&.057!&0.27 \\
\[[NII]6548+ & 1.2&-&1.4&1.4&1.8&-&2.4&3.8&2.5&2.4&2.7&2.6&2.5&2.1&1.9&2.&1.5&-&1.3&-&1.0 & 1.8\\           
\[[SII]6716 & 0.34&-&0.25&0.28&0.34&-&0.35&1.&0.37&1.&0.39&0.7&0.39&0.5&0.32&0.46&0.32&-&0.28&-&0.47&0.33 \\
\[[SII]6731 & 0.28&-&0.29&0.48&0.34&-&0.35&1.3&0.41&1.3&0.44&1.0&0.42&0.9&0.32&0.8&0.26&-&0.30&-&0.43&0.55 \\
\[[ArIII]7136 &-&-&-&-&.035&-&.039&0.5&.074&0.43&.070&0.13&.064&0.3&.042&0.23&-&-&-&-&-& - \\
\[[OII]7325 &-&-&-&-&.036&-&.054&0.2&.069&.086&.088&0.08&.099&0.17&.078&0.14&-&-&-&-&-&- \\
 $\rm H\beta$ &23*&-&50*&0.22&86*&-&220*&0.02&210*&0.05&160*&0.7 &190*&0.07&150*&0.23&63*&-&43*&-&35*&0.21  \\
 & & & & & & &&&&&&&&&&&&&&&\\
\ \Vs (\kms)&-&-&-&260&-&-&-&280&-&240&-&200&-&260&-&250&-&-&-&-&-&260  \\
\ \n0 (\cm3)&-&-&-&180&-&-&-&80&-&80&-&200&-&160&-&180&-&-&-&-&-&180  \\
\ \Ts (K)&-&-&-&-&-&-&-&-&-&-&-&-&-&-&-&-&-&-&-&-&-& - \\
\ U &-&-&-&1.2(-4)&-&-&-&1.1(-4)&-&2.2(-4)&-&1.1(-3)&-&2.35(-4)&-&2.2(-4)&-&-&-&-&-&4.2(-5)\\
\ log \Fn &-&-&-&10.&-&-&-&9.6&-&9.9&-&11.&-&10.23&-&10.25&-&-&-&-&-&  9.48 \\
\ D (cm)&-&-&-&1(17)&-&-&-&1.2(17)&-&4(17)&-&2.3(18)&-&1(17)&-&1(17)&-&-&-&-&-& 1(17) \\
&& & &  & &  && && && && && && && & \\ \hline

\end{tabular}} 

 * in $10^{-15}$ \erg

\end{table}

\begin{table}
\centerline{Table 5b}
\centerline{The spectra at P.A.=11.5$^o$ from Shield \& Filippenko (1990)}
\centerline{compared with black body models (bb)}

\tiny{

\begin{tabular}{l l llll ll lllllllllllllll}     \\ \hline

 Line  &-6" & bb6 &-4.8" & bb5 &-3.6" &bb4& -2.4" &bb3 &-1.2" &bb2 &0" (Nuc)
       & bb0 &+1.2" & bb2 &+2.4" & bb3 &+3.6" & bb4 &+4.8" & bb5 &+6" & bb6 \\
\hline\\
&& & &  & &  && && && && && && && & \\
\[[NeV]3426&-&- &-&- &-&- &-&- &-&- &0.78&- &-&- &-&- &-&- &-&- &-& - \\
\[[OII]3727& 2.7:&1.06&1.7&2.0&3.7&3.7&8.0&8.7&3.1&-&1.7&-&2.5&-&3.0&-&1.7&1.3&1.2&1.4&0.66&0.9  \\
\[[NeIII]3869+ &-&- &-&- &-&- &2.7 &1.0&1.3&- &1.1&- &1.5&- &1.9&- &-&- &-&- &-&- \\                
\[[SII]4071+   &-&- &-&- &-&- &0.96&0.05 &0.31&- &0.26&- &0.43&- &0.93&- &-&- &-&- &-&- \\
\[[OIII]4363 &-&- &-&- &-&- &-&- &-&- &0.082&- &-&- &-&- &-&- &-&- &-& - \\                         
\[HeII 4686 & - &- &-&- &-&- &0.23&0.01 &0.20&- &0.16&- &0.14&-&0.13&- &.097&.01 &-&- &-&- \\           
\[[OIII]5007 &0.37&0.25&0.74&0.8 &2.4&2.57&5.2&5.0 &6.1&-&6.0&- &5.1&-&3.0&-&1.1&0.83 &0.45&0.6&0.11&0.2 \\
\[[NI]5200+&-&-&-&-&0.10:&.01&0.10&0.01&0.16&-&0.14&-&0.12&-&0.13&-&0.075&.014&0.011&.006&-&- \\            
\[HeI 5876 &- &-&-&-&0.12:&0.13&.094&0.13&.042&-&.033&-&.047&-&.044&-&.078&.043&.097&0.13&.053 & 0.18 \\   
\[[OI]6300+ & .057!&0.1&.064!&.09&.14&0.1&.23&0.05&0.26&-&.25&-&.25&-&0.21&-&.13&0.11&.064!&.06&.057!&0.1  \\
\[[NII]6548+ & 1.2&1.&1.4&2.0&1.8&1.5&2.4&2.&2.5&-&2.7&-&2.5&-&1.9&-&1.5&0.9&1.3&1.3&1.0 & 0.8\\           
\[[SII]6716 & 0.34&0.26&0.25&0.22&0.34&0.3&0.35&0.39&0.37&-&0.39&-&0.39&-&0.32&-&0.32&0.24&0.28&0.2&0.47&0.24 \\
\[[SII]6731 & 0.28&0.41&0.29&0.37&0.34&0.2&0.35&0.39&0.41&-&0.44&-&0.42&-&0.32&-&0.26&0.39&0.30&0.29&0.43&0.36 \\
\[[ArIII]7136 &-&-&-&-&.035&0.1&.039&0.44&.074&-&.070&-&.064&-&.042&-&-&-&-&-&-& - \\
\[[OII]7325 &-&-&-&-&.036&0.18&.054&0.65&.069&-&.088&-&.099&-&.078&-&-&-&-&-&-&- \\
\ $\rm H\beta$ &23*&2(-3)&50*&.025&86*&4(-4)&220*&1.3(-4)&210*&-&160*&-&190*&-&150*&-&63*&6(-3)&43*&0.03&35*&0.06  \\
 & & & & & & &&&&&&&&&&&&&&&\\
\ \Vs (\kms)&-&230&-&200&-&250&-&100&-&-&-&-&-&-&-&-&-&200&-&200&-&230  \\
\ \n0 (\cm3)&-&200&-&20&-&20&-&100&-&-&-&-&-&-&-&-&-&200&-&200&-&200  \\
\ \Ts (K)&-&4(4)&-&5(4)&-&5(4)&-&5(4)&-&-&-&-&-&-&-&-&-&4(4)&-&5(4)&-&6(4)  \\
\ U &-&7(-3)&-&1(-2)&-&3(-3)&-&1(-3)&-&-&-&-&-&-&-&-&-&1.5(-3)&-&0.01&-& 7(-3) \\
\ log \Fn &-&-&-&-&-&-&-&-&-&-&-&-&-&-&-&-&-&-&-&-&-& - \\
\ D (cm)&-&6(17)&-&2(16)&-&3(17)&-&6(15)&-&-&-&-&-&-&-&-&-&6(17)&-&2.6(16)&-& 9(17) \\
&& & &  & &  && && && && && && && & \\ \hline

\end{tabular}} 

 * in $10^{-15}$ \erg

\end{table}

\begin{table}
\centerline{Table 6}
\centerline{Multi-cloud models}
\small{
\begin{tabular}{lll lll lll l} \\
 \hline
\hline
Line & obs& PL3 & HV1 & AV1 & HV2 & AV2 &CL1 & CL2 & AVC \\
&&&&&&&&& \\
\ [NeV] 3426 & 0.78* & 0.04 & 4.75 & 0.5& - & - & 0.04 & 0.00 & 0.02 \\
\ [OII] 3727 & 1.7* &1.1& 0.9 & 1.  &- & - & 0.32 & 4.99 & 2.64 \\
\ [NeIII] 3869+ &1.07 &0.9& 2.17 & 1.02 & 0.61 & 0.84&1.95 & 0.60 & 1.28 \\
\ [SII] 4071+ & 0.28 & 0.22 & 0.27 & 0.22 & 7(-3)& 0.17 & 0.30&0.19&0.25 \\
\ [OIII] 4363 & 0.13 & 0.05 & 1.32 & 0.16 & 8(-3) & 0.04 & 0.24& 0.01& 0.12 \\
\ HeII 4686 & 0.15 & 0.04& 0.37& 0.07& 0.06 & 0.045& 0.13&0.09&0.11 \\
\ [OIII] 5007+ &8.3 & 9.24& 11.7&9.46&4.7&8.2&16.63&1.90&9.30 \\
\ [NI] 5200+&0.12&0.10&5(-4)&0.09&0.05&0.09& 0.02 & 0.19 & 0.10 \\
\ HeI 5876 &0.06& 0.13 & 0.07&0.12&0.12&0.13&0.13&0.15&0.14 \\
\ [OI] 6300+ &0.47&0.43&0.11&0.40&0.76&0.5&0.55&0.52&0.54 \\
\ [NII] 6584+ & 4.12 & 2.6 & 0.9 & 2.45 & 2.6 & 2.6 & 1.74&5.01& 3.37 \\
\ [SII] 6716 & 0.4&0.7&8(-3)&0.6&0.02&0.5 & 0.07&1.08&0.58 \\
\ [SII] 6730 & 0.44&1.0 &0.02&0.9&0.03&0.78 &0.15&1.20&0.67 \\
\ [ArIII] 7136 & 0.07 & 0.13 & 0.27 & 0.14 & 0.4&0.2& 0.37 & 0.25 & 0.31 \\
\ [OII] 7325+ &0.14& 0.08 & 4.27& 0.11 & 0.16 & 0.1 &0.43 & 0.12 & 0.27 \\
\ \Hb (\erg) & - & 0.7 & 7(-3)&- & 0.2 & - & 4.28 & 4.28(-3) & - \\
&&&&&&&&& \\
 \Vs (\kms) & - & 200 & 1000 & - & 1000& - & - & - & -  \\
 \n0 (\cm3) &-&200& 1000 &- &50 & - & 1(5) & 1(3) & - \\
 Log(\Fn) &  - & 11 & s.d. & - & 10 & - & - & - & - \\
 Log(U)        & -  & $<$-3. & - & - & -3.2& -&-2.5 & -3.5 & - \\
 D (cm) &  - & 2.3(18) & 3(18) & - & 3(19) & - & - & - & - \\
 d/g    &  - & 1(-14)& 5(-12) & - & 2(-16) & - & - & - & - \\
&&&&&&&&& \\
 W(AV1) & - & 1 & 10  & - & - & - & - & -    & - \\
 W(AV2) & - & 1 & -   & - & 1 & - & - & -    & - \\
 W(AVC) & - & - & -   & - & - & - & 1 & 1000 & - \\
 $\rm P_{[OIII]}$(AV1) & - &0.88 & 0.12 & -&- &-&-&-&- \\
 $\rm P_{[OIII]}$(AV2) & - &0.87 & -    & -   &0.13& - &-&- \\
 $\rm P_{[OIII]}$(AVC)& - & - & -  & -   & -  & -   &0.9 & 0.1 & - \\
\hline 
\hline
\end{tabular}}
 * From Shields \& Filippenko (1990)
\end{table}

\end{document}